# Topological Properties of $\tau$-Type Organic Conductors with a Checkerboard Lattice

Toshihito Osada*

*Institute for Solid State Physics, University of Tokyo,*

*5-1-5 Kashiwanoha, Kashiwa, Chiba 277-8581, Japan.*

Although the topological phases are difficult to be realized in organic molecular crystals, we demonstrate here that they can emerge in the $\tau$-type organic layered conductors, $\tau$-(EDO-$S,S$-DMEDT-TTF)$_2$X$_{1+y}$ and $\tau$-(P-$S,S$-DMEDT-TTF)$_2$X$_{1+y}$ (X=AuBr$_2$, I$_3$, IBr$_2$), where EDO-$S,S$-DMEDT-TTF and P-$S,S$-DMEDT-TTF denote the planar donor molecules ethylenedioxy-$S,S$-dimethyl(ethylenedithio)tetrathiafulvalene and pyrazino-$S,S$-dimethyl(ethylenedithio)tetrathiafulvalene, respectively. The conducting layers of these conductors have a highly symmetric checkerboard structure, which can be regarded as a modified Mielke lattice. Because their electronic structure inherits that of the Mielke lattice, their conduction and valence bands exhibits the quadratic band touching. The contact point splits into a pair of Dirac cones under uniaxial strain which breaks C$_4$-symmetry. In $\tau$-type conductors, we can expect rather large spin-orbit coupling (SOC) as organic conductors. We show that the SOC in this case opens a topologically nontrivial gap at the band contact point, and the helical edge states exist in the gap. The actual $\tau$-type conductors could be regarded as heavily-doped topological insulators, which could exhibit finite spin Hall effect.

---

*osada@issp.u-tokyo.ac.jp

# I. INTRODUCTION: TOPOLOGY OF ORGANIC CONDUCTORS

It is desirable that topological phases can be realized in organic molecular crystals with high designability. However, the search for topological states in organic crystals has met with some challenges. Generally, organic molecular crystals have tight-binding band structures with narrow band width, making it difficult to cause band inversion. Moreover, organic molecules consist of light elements with correspondingly small SOC.

So far, only a few molecular crystals have been studied as topological materials in the field of organic conductors. Under high pressure $P > 1.2$ GPa suppressing the charge order transition, $\alpha$-(BEDT-TTF)$_2$I$_3$, where BEDT-TTF denotes bis(ethylenedithio)-tetrathiafulvalene, was revealed to be a two dimensional (2D) Dirac fermion system [1]. This fact was first pointed out by a tight-binding study [2, 3], and later confirmed by the first principles density functional theory (DFT) calculations [4]. Experimentally, the realization of Dirac fermions was indirectly suggested by magnetotransport [5-10], specific heat [11], thermoelectric power [12], and NMR measurements [13]. Based on this Dirac fermion state, the Chern insulator state and the $Z_2$ topological insulator state were investigated by introducing magnetic modulation and SOC, respectively [14, 15]. In the single component molecular conductor Pd(dddt)$_2$, where dddt denotes 5,6-dihydro-1,4-dithiin-2,3-dithiolate, a three-dimensional (3D) nodal-line Dirac semimetal state was suggested under high pressure ($P > 12.8$ GPa) by DFT calculations and transport measurements using the diamond anvil cell [16-18]. Recently, another single component conductor Pt(dmdt)$_2$, where dmdt denotes dimethyltetrathiafulvalenedithiolate, has been suggested to be a nodal-line semimetal at ambient pressure [19]. In addition, a DFT calculation has suggested that (BEDT-TTF)Ag$_4$(CN)$_5$, in which BEDT-TTF molecules form a distorted diamond lattice, is a nodal-line Dirac semimetal [20]. Because (BEDT-TTF)Ag$_4$(CN)$_5$ is an antiferromagnetic Mott insulator at ambient pressure, high pressure experiments have been tried to suppress the insulating state.

## II. $\tau$-TYPE ORGANIC CONDUCTORS

In this paper, we propose that the $\tau$-type organic conductors $\tau$-(EDO-$S$,$S$-DMEDT-TTF)$_2$X$_{1+y}$ and $\tau$-(P-$S$,$S$-DMEDT-TTF)$_2$X$_{1+y}$ (X=AuBr$_2$, I$_3$, IBr$_2$) are new topological organic molecular crystals, and discuss their topological properties. These $\tau$-type organic conductors have a unique crystal structure [21]. They are layered conductors in which a conducting layer and an anion layer stack alternately. In each conducting layer, donor molecules form a square lattice, and anion molecules (X) are arranged on it with a checkerboard pattern. It is characteristic to $\tau$-type conductors that each conducting layer contains anion molecules. Meanwhile, in each anion layer, anion molecules randomly occupy about 75~87.5% of anion sites ($y = 0.75 \sim 0.875$), so that $\tau$-type conductors are nonstoichiometric.

A tight-binding band calculation on the conducting layer shows that the conduction and valence bands exhibit quadratic band touching at the corner of the square Brillouin zone. Since $(1-y)/2$ of the conduction band are stoichiometrically occupied by electrons, the Fermi level is located in the conduction band, resulting in a star-shaped electron Fermi surface [21]. This fact was also confirmed by the DFT calculation [22]. The Fermi surface has been investigated by observing Shubnikov-de Haas oscillation of magnetoresistance in these salts [23-25].

The crystal structure of a conducting layer of the $\tau$-type organic conductors $\tau$-(EDO-$S$,$S$-DMEDT-TTF)$_2$X$_{1+y}$ and $\tau$-(P-$S$,$S$-DMEDT-TTF)$_2$X$_{1+y}$ is schematically shown in Fig. 1(a). Donor molecules form a square lattice, in which the unit cell (indicated here by a pale square) contains two molecular sites, A and B, with different molecular orientations. The long axis of planar molecules is normal to the conducting plane. The nearest neighbor (NN) electron hopping between A and B has a transfer integral $t_1$. The next nearest neighbor (NNN) hopping between A and A (B and B) has two types of transfer integrals, $t_2$ and $t_3$, for the face-to-face and side-by-side hopping, respectively. Because most of the probability density of the highest occupied molecular orbital (HOMO) exists in both sides of a planar molecule, the face-to-face contact is much larger than the side-by-side contact. The anion site, which is located in the center of the face-to-face contact, might affect $t_2$, whereas the anion orbital forms a closed shell and does not contribute to the band structure.

The lattice structure of the $\tau$-type organic conductors can be regarded as a modified Mielke lattice. The Mielke lattice has a checkerboard structure as shown in Fig. 1(b), in which the face-to-face NNN hopping integral ($t_2$) takes the same value ($t$) as the NN hopping integral ($t_1$), and the side-by-side NNN hopping integral ($t_3$) is zero. The Mielke lattice has a flat band and a cosine band which exhibit quadratic band touching. The emergence of flat-band ferromagnetism was discussed using the Hubbard model on the Mielke lattice [26, 27]. Recently, topological many-body states have been discussed on the checkerboard lattice [28, 29]. The $\tau$-type organic conductors might inherit the topological features of the Mielke lattice. In fact, quadratic band touching is a common feature of these two systems.

Recently, Winter, Riedl, and Valentí pointed out the significance of SOC in layered organic molecular crystals based on DFT calculation [30]. In the tight binding model, SOC is represented by the additional spin-dependent complex hopping energy $(i/2)\boldsymbol{\lambda}_{ij} \cdot \mathbf{s}$, where $\mathbf{s}$ and $\boldsymbol{\lambda}_{ij}$ denote the electron spin, and the SOC strength associated with the hopping from the *j*-th molecular site to the *i*-th one, respectively. They showed that $\boldsymbol{\lambda}_{ij}$ is normal to the layer, so that the SOC hopping is written as $\pm(i/2)|\boldsymbol{\lambda}_{ij}|s_z$, where $s_z(=\pm 1)$ is the normal component of $\mathbf{s}$. They also showed that $|\boldsymbol{\lambda}_{ij}|$ is maximal when the dihedral angle between the *i*-th and *j*-th molecules is $\pi/2$. They estimated $|\boldsymbol{\lambda}_{ij}|$ in BEDT-TTF and BETS families of organic conductors to be in the order of 1 ~ 2 meV and 5 ~ 10 meV, respectively, where BETS denotes bis(ethylenedithio)tetraselenafulvalene. This value of SOC is small but can cause observable effects, especially at the band contact point.

In $\tau$-type organic conductors, the dihedral angle between A and B is $\pi/2$, which gives the maximum SOC. Moreover, the anions containing heavy elements are located on the conduction layer in a checkerboard pattern, so that an electron hopping between NN molecules should feel strong transverse in-plain potential gradient generated by anions on one side of the hopping path. Therefore, we can expect rather strong SOC in $\tau$-type organic conductors.

### III. ORGANIC VERSION OF KANE-MELE MODEL

We consider a tight-binding model for a single 2D conduction layer of the $\tau$-type organic conductor taking into account SOC. The SOC of the 2D molecular layer depends only on $s_z$, the normal component of spin, as mentioned in the previous section. Since $s_z$ is a good quantum number, the Hamiltonian matrix is decoupled into two spin sectors as $H(\mathbf{k}) = H(\mathbf{k}, s_z = +1) \oplus H(\mathbf{k}, s_z = -1)$. Here, $H(\mathbf{k}, s_z)$ is a $2 \times 2$ matrix, whose bases are the Bloch sums constructed from the HOMOs of the molecules A and B with a spin $s_z$.

$$H(\mathbf{k}, s_z) = \begin{pmatrix} H_{AA}(\mathbf{k}) & H_{AB}(\mathbf{k}, s_z) \\ H_{AB}(\mathbf{k}, s_z)^* & H_{BB}(\mathbf{k}) \end{pmatrix},$$

$$H_{AB}(\mathbf{k}, s_z) = (t_1 - i\lambda s_z)e^{-ik_x\frac{a}{2}-ik_y\frac{a}{2}} + (t_1 + i\lambda s_z)e^{-ik_x\frac{a}{2}+ik_y\frac{a}{2}}$$
$$+ (t_1 + i\lambda s_z)e^{+ik_x\frac{a}{2}-ik_y\frac{a}{2}} + (t_1 - i\lambda s_z)e^{+ik_x\frac{a}{2}+ik_y\frac{a}{2}},$$

$$H_{AA}(\mathbf{k}) = t_{2A}(e^{-ik_x a} + e^{+ik_x a}) + t_3(e^{-ik_y a} + e^{+ik_y a}),$$

$$H_{BB}(\mathbf{k}) = t_{2B}(e^{-ik_y a} + e^{+ik_y a}) + t_3(e^{-ik_x a} + e^{+ik_x a}). \tag{1}$$

The energy of the HOMOs of the donor molecules are taken as zero. Following the pioneering work of Kane and Mele [31], we introduce SOC into the inter-site matrix element $H_{AB}(\mathbf{k}, s_z)$ as additional NN hopping terms with spin-dependent complex hopping energy $\pm i\lambda s_z$. Here, $\lambda$ represents the strength of SOC, and the double sign becomes positive (negative) when an anion is located in the right (left) side of the hopping path. In Fig. 1(a), the NN hopping along a solid (dashed) circle surrounding an anion X takes a positive (negative) sign. Note that the NNN hopping has no additional SOC term because the both sides of the hopping path is symmetric. In addition to the SOC, we introduce anisotropy of the NNN transfer $t_2$ to discuss the effects of uniaxial strain. We set face-to-face transfer between A and A (B and B) sites as $t_{2A}$ ($t_{2B}$). Normally, $t_{2A}$ and $t_{2B}$ are equal, whereas they takes different value under the uniaxial strain in the $x$ ($y$) direction.

Comparing with the result of the DFT calculation [22], we employ $t_1 \equiv t$, $t_{2A} = t_{2B} \equiv t_2 = 0.13\,t$ and $t_3 = -0.07\,t$ with $t = 0.16$ eV for $\tau$-(EDO-S,S-DMEDT-TTF)$_2$(AuBr$_2$)$_{1+y}$. The tight binding model used for fitting of DFT results in Ref. 22 (Eq. (5)) differs from the present model by the phase factor $\exp(ik_x a/2 + ik_y a/2)$, and it has two NN transfer integrals with opposite sign. Considering these facts, we have chosen the above transfer integrals. Note that the **k**-space of the present model is shifted by $(\pi/a, \pi/a)$ from that of Ref. 22. The M-point of the present model corresponds to the Γ-point in Ref. 22. Also note that we have chosen a negative value of $t_3$ according to Ref. 22, although most of previous tight-binding calculations used positive $t_3$ [21, 25]. The topological features discussed below are not affected by the sign of $t_3$. The large ratio of $t_1$ to $t_2$ and the existence of finite $t_3$ are differences of the present model from the Mielke lattice. The Hamiltonian is easily diagonalized, and the following dispersion is obtained for the isotropic case ($t_{2A} = t_{2B} \equiv t_2$).

$$E_\pm(\mathbf{k}) = (t_2 + t_3)(\cos ak_x + \cos ak_y)$$
$$\pm \sqrt{(t_2 - t_3)^2(\cos ak_x - \cos ak_y)^2 + 16t_1^2 \cos^2\left(\frac{ak_x}{2}\right)\cos^2\left(\frac{ak_y}{2}\right) + 16\lambda^2 \sin^2\left(\frac{ak_x}{2}\right)\sin^2\left(\frac{ak_y}{2}\right)}. \quad (2)$$

Here, "+" and "−" of the double sign correspond to the conduction band $E_2(\mathbf{k})$ and the valence band $E_1(\mathbf{k})$, respectively. These bands have twofold spin degeneracy.

The present model is an organic version of the Kane-Mele model for graphene [31]. In a previous work, we discussed the effect of SOC in an organic Dirac fermion system α-(BEDT-TTF)$_2$I$_3$ in a similar way [15]. In contrast to the original Kane-Mele model, which assumes unrealistic NNN hopping in the honeycomb lattice, the present model assumes no additional hopping in the checkerboard lattice.

In the absence of anisotropy and SOC ($t_{2A} = t_{2B}$ and $\lambda = 0$), the above model gives fourfold-symmetric conduction and valence bands with quadratic band touching at the corner of square Brillouin zone (M point) as shown in Fig. 2(a). The Fermi level is located in the conduction band because of $y = 0.75 \sim 0.875$, forming a fourfold-symmetric star-shaped Fermi surface of electrons. The "fins" of the valence band $E_1(\mathbf{k})$ along the M-X and M-Y lines mainly consist of A-site and B-site states, respectively. Meanwhile, in the conduction band $E_2(\mathbf{k})$, the M-X and M-Y fins mainly consists of B-site and A-site states, respectively. The system has the inversion symmetry, so that the Berry curvature is zero over the Brillouin zone except at the band contact point.

## IV. DIRAC FERMIONS UNDER UNIAXIAL STRAIN

First, we consider the effect of uniaxial strain on the crystal lattice of our $\tau$-type conductor. This strain breaks the $C_4$ symmetry of this lattice. When we compress the crystal in the $x$ direction, the most affected transfer is the face-to-face NNN transfer $t_{2A}$. This is because the length of the NNN bond between A and A is the most decreased under this strain. Therefore, the effect of uniaxial strain can be simulated by introducing differences only between $t_{2A}$ and $t_{2B}$, even though other transfers are changed under the strain. Figure 2(b) shows the band dispersion for the anisotropic case of $t_{2A} > t_{2B}$ ($t_{2A} = 1.2t_2$ and $t_{2B} = 0.8t_2$). In this situation, a pair of Dirac cones appears, breaking their merging at the quadratic band touching point. The valence and conduction bands at the original touching point M now consist of A-site and B-site states, respectively. This is a topological transition from a zero-gap state with quadratic band touching (2D Luttinger semimetal) to one with twin Dirac cones (2D Dirac semimetal). The strain-induced topological transition from the Luttinger to Dirac semimetal was similarly discussed in the Shastry-Sutherland lattice [32].

## V. TOPOLOGICAL INSULATOR STATE WITH FINITE SPIN-ORBIT COUPLING

Next, we discuss the effect of finite SOC ($\lambda \neq 0$) in $\tau$-type organic conductors for both cases with zero and with finite uniaxial strain. As mentioned above, we can expect rather strong SOC in $\tau$-type conductors. The SOC in the NN hopping opens a finite gap at the band contact point for the both cases of $t_{2A} = t_{2B}$ and $t_{2A} > t_{2B}$ as shown in the upper panels of Fig. 3(a) and 3(b), respectively. In the figure, $\lambda = 0.01t$ is used for the SOI strength. In the followings, we see that this gapped state is a topologically non-trivial, although the Fermi energy is located in the conduction band.

Fu and Kane found a simple criterion for a topological insulator with space inversion symmetry [33]. On the $n$-th spin-degenerated band, an inversion operator has a parity eigenvalue $\xi_n(\mathbf{k}_{\text{TRIM}})$ of $+1$ or $-1$ at time-reversal invariant momenta (TRIMs), which satisfy $-\mathbf{k}_{\text{TRIM}} = \mathbf{k}_{\text{TRIM}} + \mathbf{G}$ with a reciprocal lattice vector $\mathbf{G}$. If the product of $\xi_n(\mathbf{k}_{\text{TRIM}})$ for all TRIMs in the Brillouin zone and for all band below the energy gap becomes $-1$, the system is a topological insulator when the Fermi energy is located in the gap. In the $\tau$-type conductor, the crystal lattice has inversion symmetry with an inversion center at the point labeled "c" in Fig. 1(a) even in the case of $t_{2A} \neq t_{2B}$ and/or $\lambda \neq 0$. In its conduction and valence bands, up-spin ($s_z = +1$) and down-spin ($s_z = -1$) subbands are degenerated. There exist four TRIMs in the square 2D Brillouin zone at $\mathbf{k}_{\text{TRIM}} = (0, 0)$, $(\pi/a, 0)$, $(0, \pi/a)$, and $(\pi/a, \pi/a)$, at which the parity eigenvalues of the valence band are $+1$, $+1$, $+1$, and $-1$, respectively. Since the parity product is equal to $-1$, we can conclude that the gapped $\tau$-type conductor is a non-trivial topological insulator when the Fermi level is located in the energy gap. Note that the parity product is unchanged, even if we employ other inversion centers giving different parity values.

In fact, the gapped state has a finite spin-dependent Berry curvature. Since $H(\mathbf{k})$ is decoupled into two spin sectors, we can calculate the Berry curvature separately for each spin-subband. The $z$-component of spin-dependent Berry curvature

$\Omega_n(\mathbf{k}, s_z)$ at **k**-point of the *n*-th band (*n* = 1, 2) is given by

$$[\Omega_n(\mathbf{k}, s_z)]_z = i \sum_{m(\neq n)} \frac{\langle u_{n\mathbf{k}} | \partial H(\mathbf{k}, s_z)/\partial k_x | u_{m\mathbf{k}} \rangle \langle u_{m\mathbf{k}} | \partial H(\mathbf{k}, s_z)/\partial k_y | u_{n\mathbf{k}} \rangle - c.c.}{\{E_m(\mathbf{k}) - E_n(\mathbf{k})\}^2}. \quad (3)$$

Here, $E_n(\mathbf{k})$ and $|u_{n\mathbf{k}}\rangle$ denote the eigenenergy and the two-component eigenvector of $H(\mathbf{k}, s_z)$, respectively. "c.c." denotes the complex conjugate of the first term of the numerator. The contribution of each $s_z = \pm 1$ spin-subband of the *n*-th band $E_n(\mathbf{k})$ to the off-diagonal conductivity $\sigma_{xy}$ is given by the following formula at zero temperature and zero magnetic field.

$$\sigma_{xy}^{(n)}(s_z) = -\frac{e^2}{h} \left[ \frac{1}{2\pi} \iint_{E_n(\mathbf{k}) < E_F} [\Omega_n(\mathbf{k}, s_z)]_z dk_x dk_y \right]. \quad (4)$$

Here, $E_F$ indicates the Fermi level, so that the above integral is taken inside the Fermi surface. The lower panels of Fig. 3(a) and 3(b) show the *z*-component of the Berry curvature of both spin-subbands of the valence band $E_1(\mathbf{k})$. Each Berry curvature exhibits four peaks in the case of $t_{2A} = t_{2B}$ and two peaks in the case of $t_{2A} > t_{2B}$ with the same sign. However, the sign is opposite for up-spin and down-spin subbands. This means that the carriers with different spin obtain opposite directions of anomalous velocity under the external in-plane electric field. This causes no charge current but finite spin current in the direction perpendicular to the electric field, namely, the spin Hall effect. In fact, the spin Hall conductivity of the valence band, $\sigma_{xy}^{(1)}(+1) - \sigma_{xy}^{(1)}(-1)$, becomes finite.

When the Fermi level is located in the energy gap, the system exhibits the quantum spin Hall effect. According to the theory by Thouless, Kohmoto, Nightingale, and den Nijs (TKNN) [34, 35], the off-diagonal conductivity $\sigma_{xy}^{(1)}(s_z = \pm 1)$ of occupied spin-subband is quantized to $-N_{\pm}e^2/h$, where $N_{\pm}$ is an integer. Since $N_- = -N_+$, the spin Hall conductivity $\sigma_{xy}^{(1)}(+1) - \sigma_{xy}^{(1)}(-1)$ is quantized to $-2N_+e^2/h$.

In addition, we investigate the edge state of a 2D nanoribbon of the $\tau$-type conductor parallel to the $x$-axis, which has two types of edges, the AA-edge and BB-edge as shown in Fig. 1(a). At the AA-edge, the molecular plane is normal to the crystal surface, whereas it is parallel to the surface at the BB-edge. By using a similar method to that used in analyzing $\alpha$-(BEDT-TTF)$_2$I$_3$ [36], we calculated the $k_x$-dispersion of the energy spectra as shown in Fig. 4(a) and 4(b) for the cases of $t_{2A} = t_{2B}$ and $t_{2A} > t_{2B}$, respectively. The isolated branches, which appear in the bulk gap, are the edge states with up-spin or down-spin localized around the AA-edge or BB-edge, as indicated by labels. We can see that a pair of edge states with opposite spin and group velocity, called the helical edge state, appears both along the AA-edge and BB-edge. The helical edge state carries no charge current but finite spin current along the sample edge. The appearance of a metallic helical edge state is one of the most characteristic features of topological insulators when the Fermi level is in the energy gap. In the dispersion of Fig. 4, the region labeled as "A-like" ("B-like") mainly consists of A-site (B-site) states. We can see that the helical edge state connects the regions with the same dominant site of the conduction and valence bands. For example, the AA-edge state connects A-like regions of the conduction and valence bands. This feature reflects the band inversion.

## VI. POSSIBILITY OF SPIN HALL EFFECT

The actual $\tau$-type conductors are regarded as heavily doped topological insulators where the Fermi level is in the conduction band. The Fermi energy is estimated as 40 ~60 meV from the band contact point. The SOC gap and the helical edge state could be detected by angle-resolved photoemission spectroscopy (ARPES) measurement. Here, we consider the possibility to observe the topological transport phenomena, namely the spin Hall effect in the actual $\tau$-type conductors.

Let us estimate the spin Hall conductivity in $\tau$-(EDO-*S*,*S*-DMEDT-TTF)$_2$(AuBr$_2$)$_{1+y}$. We assume a SOC strength of $\lambda = 0.01t = 1.6$ meV so as to be in the same order as BEDT-TTF family, although larger SOC could be expected in the $\tau$-type conductors. This opens a SOC gap of 12.8 meV. The total spin Hall conductivity for both the valence and conduction bands is given by $\sigma_{xy}^{\text{spin}} = \{\sigma_{xy}^{(1)}(+1) - \sigma_{xy}^{(1)}(-1)\} + \{\sigma_{xy}^{(2)}(+1) - \sigma_{xy}^{(2)}(-1)\}$. Since $\sigma_{xy}^{(1)}(+1) - \sigma_{xy}^{(1)}(-1)$ is quantized to $-2 \times (-1)e^2/h$ in the occupied valence band, $\sigma_{xy}^{\text{spin}}$ is calculated by the following formula.

$$\sigma_{xy}^{\text{spin}} = -\frac{2e^2}{h}\left[-1 + \frac{1}{2\pi}\iint_{E_n(\mathbf{k})<E_F} [\mathbf{\Omega}_2(\mathbf{k}, s_z = +1)]_z \, dk_x dk_y \right]. \tag{5}$$

Figure 5(a) shows the Fermi surfaces for several anion contents and **k**-dependence of $a^2[\mathbf{\Omega}_2(\mathbf{k}, s_z = +1)]_z$ for the conduction band, where *a* is a lattice constant. By taking an integral inside the Fermi surface, we obtain the filling dependence of the spin Hall conductivity as shown in Fig. 5(b). In addition to the case of $\lambda = 0.01t$, the spin Hall conductivity for several other SOC values are plotted. The band filling is represented both by the anion content *y* and by the cross section of the Fermi surface (normalized by the Brillouin zone area $S_{\text{BZ}}$). The Fermi level reaches the van Hove singularity at the filling indicated by an arrow, resulting in the slight change of slope. From Fig. 5(b), the spin Hall conductivity $\sigma_{xy}^{\text{spin}}$ of $\tau$-(EDO-*S*,*S*-DMEDT-TTF)$_2$(AuBr$_2$)$_{1+y}$ with anion content of $0.75 < y < 0.875$ (indicated by the shaded region) is estimated as $0.15\, e^2/h$ ~ $0.36\, e^2/h$ for $\lambda = 0.01t$. Since $\sigma_{xy}^{\text{spin}}$ is still about 10% of the quantized value, the spin Hall effect could survive even in the actual $\tau$-type conductors.

The spin Hall effect is usually observed by the nonlocal resistance measurement. The nonlocal resistance is proportional to $(\sigma_{xy}^{\text{spin}})^2 / (\sigma_{xx})^3 \times \exp(L/l^{\text{spin}})$ in its measurement configuration, where $\sigma_{xx}$, $L$, and $l^{\text{spin}}$ are the 2D normal conductivity due to Ohmic transport, the electrode distance, and the spin diffusion length due to spin inversion scattering, respectively [37]. Therefore, in order to obtain a sufficient signal, it is necessary to use a sample with $\sigma_{xx}$ as small as possible and $L$ as short as possible. Using the results of the previous experiment [23], $\sigma_{xx}$ is estimated to be much less than $e^2/h$, and it depends on temperature. In graphene, $l^{\text{spin}}$ is estimated to be in the order of several micrometers. Therefore, the measurement should be performed at a proper temperature giving lower $\sigma_{xx}$, using a micro-fabricated sample.

## VII. TOPOLOGICAL FEATURES OF THE CHECKERBOARD MIELKE LATTICE

The checkerboard Mielke lattice is obtained by choosing a special set of transfer integrals, namely, $t_1 = t$, $t_2 (\equiv t_{2A} = t_{2B}) = t$, and $t_3 = 0$ in the present model. Therefore, the Mielke lattice shows the topological features of the $\tau$-type conductors described above: the quadratic band touching, the topological transition to the Dirac fermion system under uniaxial strain, and the topologically nontrivial gap at the band contact point under SOC.

Here, we see the other features characteristic to the Mielke lattice. Figure 6(a) exhibits the band dispersion of the Mielke lattice with no SOC. The conduction band is the cosine band, as for a square lattice, and the valence band is a dispersion-less flat band. These band show the quadratic band touching at the M-point. The fin-like structures in the band dispersion of the $\tau$-type conductors are the remnants of the flat band feature of the Mielke lattice. In fact, in the $\tau$-type conductors, the weak band dispersion along the boundaries of Bulliouin zone (M-X and M-Y lines) can be regarded as a partially almost-flat band, so that the flat-band physics was discussed there [38].

Figure 6(b) exhibits the energy spectrum of the nanoribbon of the Mielke lattice with no SOC. We find an edge state localized around the BB edge, where B sites have no transfer contact with each other on the edge. As the ratio of $t_1$ to $t_2$ increases, this edge state approaches the valence band, and it matches the valence band in the $\tau$-type conductors. If we introduce the additional SOC terms in the Mielke lattice in the same way as the $\tau$-type, this BB-edge state exhibits spin splitting and becomes a helical edge state connecting the conduction and the valence band. Moreover, another helical edge state appears along the AA-edge.

## VIII. CONCLUSIONS

We have described the topological features of the $\tau$-type organic conductors, $\tau$-(EDO-*S*,*S*-DMEDT-TTF)$_2$X$_{1+y}$ and $\tau$-(P-*S*,*S*-DMEDT-TTF)$_2$X$_{1+y}$ (X=AuBr$_2$, I$_3$, IBr$_2$), of which the checkerboard lattice structure has the form of a modified Mielke lattice. The conduction and valence bands exhibits quadratic band touching inherited from the Mielke lattice. When the C$_4$ symmetry is broken by uniaxial strain, the band touching point splits into a pair of Dirac cones. The SOC, which is expected to be rather strong in this system, opens an energy gap at the band contact point regardless of uniaxial strain. This gapped state is topologically non-trivial, and a helical edge state appears in the gap. Finite spin Hall conductivity survives even in the actual $\tau$-type conductor in which electrons are heavily doped, although its experimental detection needs microfabrication of samples.

ACKNOWLEDGEMENTS

The author thanks Dr. Takako Konoike for introducing $\tau$-type organic conductors. He also thank Prof. Hideo Aoki for helpful discussion and informing their relating works. This work was partially supported by JSPS KAKENHI Grant Number JP16H03999.

[1] K. Kajita, Y. Nishio, N. Tajima, Y. Suzumura, and A. Kobayashi, J. Phys. Soc. Jpn. **83**, 072002 (2014).

[2] S. Katayama, A. Kobayashi, and Y. Suzumura, J. Phys. Soc. Jpn. **75,** 054705 (2006).

[3] A. Kobayashi, S. Katayama, Y. Suzumura, and H. Fukuyama, J. Phys. Soc. Jpn. **76**, 034711 (2007).

[4] H. Kino and T. Miyazaki, J. Phys. Soc. Jpn. **75,** 034704 (2006).

[5] T. Osada, J. Phys. Soc. Jpn. **77,** 084711 (2008).

[6] N. Tajima, S. Sugawara, R. Kato, Y. Nishio, and K. Kajita, Phys. Rev. Lett. **102**, 176403 (2009).

[7] T. Osada, J. Phys. Soc. Jpn. **80,** 033708 (2011).

[8] M. Sato, K. Miura, S. Endo, S. Sugawara, N. Tajima, K. Murata, Y. Nishio, and K. Kajita, J. Phys. Soc. Jpn. **80,** 023706 (2011).

[9] S. Sugawara, M. Tamura, N. Tajima, R. Kato, M. Sato, Y. Nishio, and K. Kajita, J. Phys. Soc. Jpn. **79,** 113704 (2010).

[10] M. Monteverde, M. O. Goerbig, P. Auban-Senzier, F. Navarin, H. Henck, C. R. Pasquier, C. Meziere, and P. Batail, Phys. Rev. B **87**, 245110 (2013).

[11] T. Konoike, K. Uchida, and T. Osada, J. Phys. Soc. Jpn. **81**, 043601 (2012).

[12] T. Konoike, M. Sato, K. Uchida, and T. Osada, J. Phys. Soc. Jpn. **82**, 073601 (2013).

[13] M. Hirata, K. Ishikawa, K. Miyagawa, M. Tamura, C. Berthier, D. Basko, A. Kobayashi, G. Matsuno, and K. Kanoda, Nat. Comms. **7,** 12666 (2016).

[14] T. Osada, J. Phys. Soc. Jpn. **86**, 123702 (2017).

[15] T. Osada. J. Phys. Soc. Jpn. **87,** 075002 (2018).

[16] R. Kato, H. Cui, T. Tsumuraya, T. Miyazaki, and Y. Suzumura, J. Am. Chem. Soc. **139,** 1770 (2017).

[17] R. Kato and Y. Suzumura, J. Phys. Soc. Jpn. **86**, 064705 (2017).

[18] Z. Liu, H. Wang, Z. F. Wang, J. Yang, and F. Liu, Phys. Rev. B **97**, 155138 (2018).

[19] B. Zhou, S. Ishibashi, T. Ishii, T. Sekine, R. Takehara, K. Miyagawa, K. Kanoda, E. Nishibori, and A. Kobayashi, Chem. Commun. **55**, 3327 (2019).

[20] Y. Shimizu, A. Otsuka, M. Maesato, M. Tsuchiizu, A. Nakao, H. Yamochi, T. Hiramatsu, Y. Yoshida, and G. Saito, Phys. Rev. B **99**, 174417 (2019).

[21] G. C. Papavassiliou, D. J. Lagouvardos, J. S. Zambounis, A. Terzis, C. P. Raptopoulou, Keizo Murata, N. Shirakawa, L. Ducasse, and P. Delhaes, Mol. Cryst. Liq. Cryst. **285**, 83 (1996).

[22] H. Aizawa, K. Kuroki, H. Yoshino, G. A. Mousdis, G. C. Papavassiliou, and K. Murata, J. Phys. Soc. Jpn. **83**, 104705 (2014).

[23] T. Konoike, K. Iwashita, H. Yoshino, K. Murata, T. Sasaki, and G. C. Papavassiliou, Phys. Rev. B **66**, 245308 (2002).

[24] G. C. Papavassiliou, G. A. Mousdis, G. C. Anyfantis, K. Murata, L. Li, H. Yoshino, H. Tajima, T. Konoike, J. S. Brooks, D. Graf, and E. S. Choi, Zeitschrift für Naturforschung A **59**, 952 (2004).

[25] H. Yoshino, K. Murata, T. Nakanishi, L. Li, E. S. Choi, D. Graf, J. S. Brooks, Y. Nogami, and G. C. Papavassiliou, J. Phys. Soc. Jpn. **74,** 417 (2005).

[26] A. Mielke, J. Phys. A: Math. Gen. **24**, L73 (1991).

[27] A. Mielke, J. Phys. A: Math. Gen. **24**, 3311 (1991).

[28] E. J. Bergholtz and Z. Liu, Int. J. Mod. Phys. B **27**, 1330017 (2013).

[29] W. Li, D. N. Sheng, C. S. Ting, and Y. Chen, Phys. Rev. B **90**, 081102(R) (2014).

[30] S. M. Winter, K. Riedl, and R. Valentí, Phys. Rev. B **95**, 060404 (2017).

[31] C. L. Kane and E. J. Mele, Phys. Rev. Lett. **95**, 226801 (2005).

[32] T. Kariyado and Y. Hatsugai, Phys. Rev. B **88**, 245126 (2013).

[33] L. Fu and C. L. Kane, Phys. Rev. B **76**, 045302 (2007).

[34] D. J. Thouless, M. Kohmoto, M. P. Nightingale, and M. den Nijs, Phys. Rev. Lett. **49**, 405 (1982).

[35] M. Kohmoto, Ann. Phys. **160**, 343 (1985).

[36] Y. Hasegawa and K. Kishigi, J. Phys. Soc. Jpn. **80**, 054707 (2011).

[37] D. A. Abanin, A. V. Shytov, L. S. Levitov, and B. I. Halperin, Phys. Rev. B **79**, 035304 (2009).

[38] R. Arita, K. Kuroki, and H. Aoki, Phys. Rev. B **61**, 3207 (2000).

**Figure 1** (T. Osada)

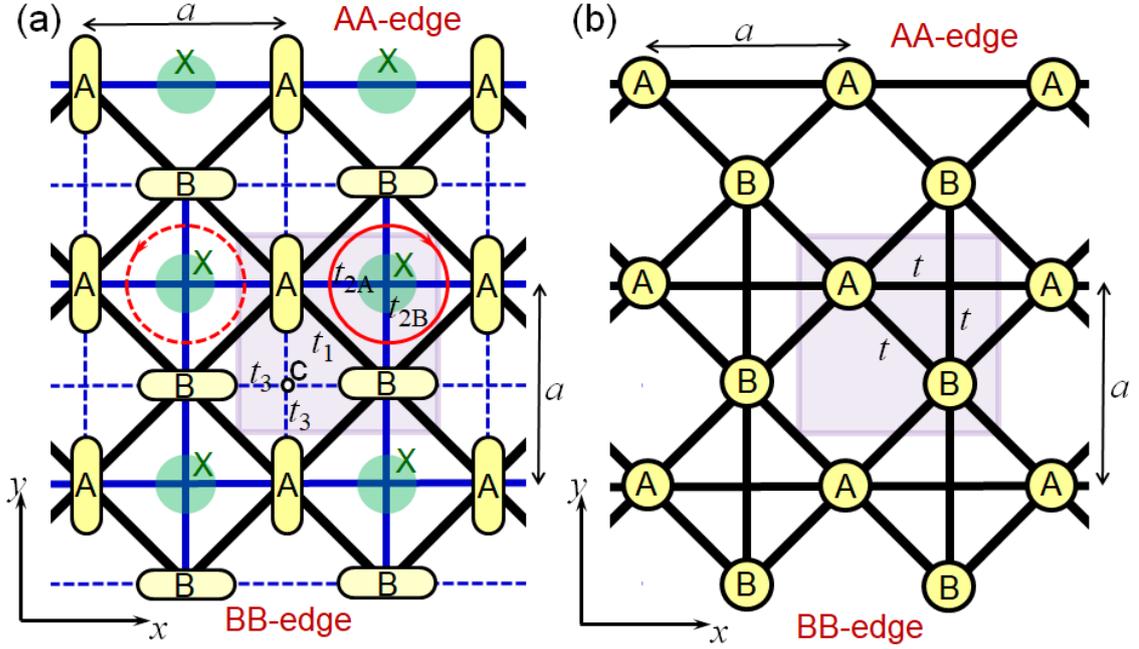

**FIG. 1.** (color online)

(a) Schematic of lattice structure of a conduction layer of a $\tau$-type organic conductor. The unit cell (indicated by a pale square) contains two molecular sites labeled A and B and an anion site labeled X. The inversion center used for the calculation of parity is labeled c. The transfer integrals, $t_1$, $t_{2A}$, $t_{2B}$, and $t_3$ are also indicated. The solid and dashed circles indicate the sign of the additional SOI terms. The AA-edge and BB-edge of the nanoribbon parallel to $x$ are exhibited at the top and bottom of the diagram. (b) Schematic of a checkerboard Mielke lattice. There are two sites, A and B in a unit cell, and transfer integrals take the same value of $t$.

**Figure 2** (T. Osada)

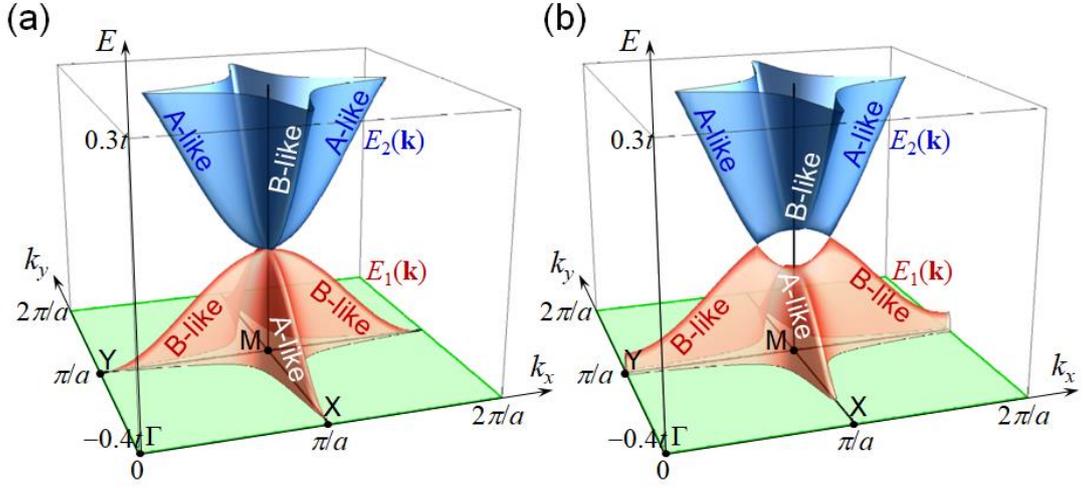

**FIG. 2.** (color online)

Dispersion of the valence band $E_1(\mathbf{k})$ and conduction band $E_2(\mathbf{k})$ of a $\tau$-type conductor for (a) the isotropic case with no strain ($t_1 \equiv t$, $t_{2A} = t_{2B} = 0.13\,t$, $t_3 = -0.07\,t$), and (b) the anisotropic case under uniaxial strain ($t_1 \equiv t$, $t_{2A} = 0.13\,t \times 1.2$, $t_{2B} = 0.13\,t \times 0.8$, $t_3 = -0.07\,t$). The fin labeled as "A-like" ("B-like") mainly consists of A-site (B-site) states.

**Figure 3** (T. Osada)

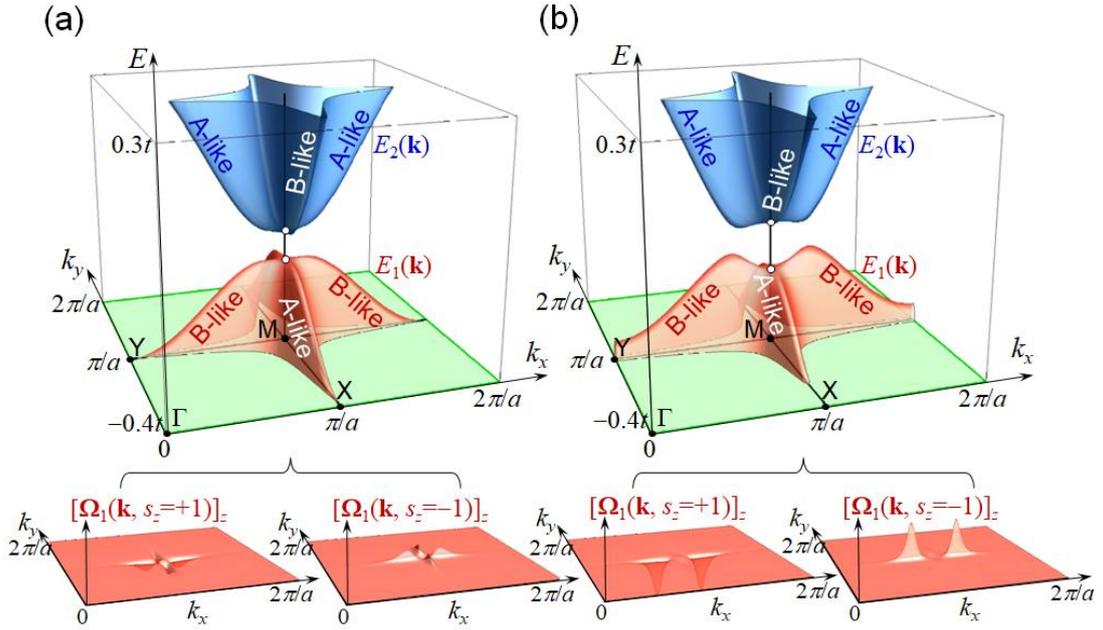

**FIG. 3.** (color online)

Dispersion of the valence band $E_1(\mathbf{k})$ and conduction band $E_2(\mathbf{k})$ of a $\tau$-type conductor with a finite SOC of $\lambda = 0.01t$. The lower panels show the $z$-component of the Berry curvature of both spin-subbands of the valence band. (a) The isotropic case with no strain ($t_1 \equiv t$, $t_{2A} = t_{2B} = 0.13\,t$, $t_3 = -0.07\,t$). (b) The anisotropic case with finite uniaxial strain ($t_1 \equiv t$, $t_{2A} = 0.13\,t \times 1.2$, $t_{2B} = 0.13\,t \times 0.8$, $t_3 = -0.07\,t$).

**Figure 4** (T. Osada)

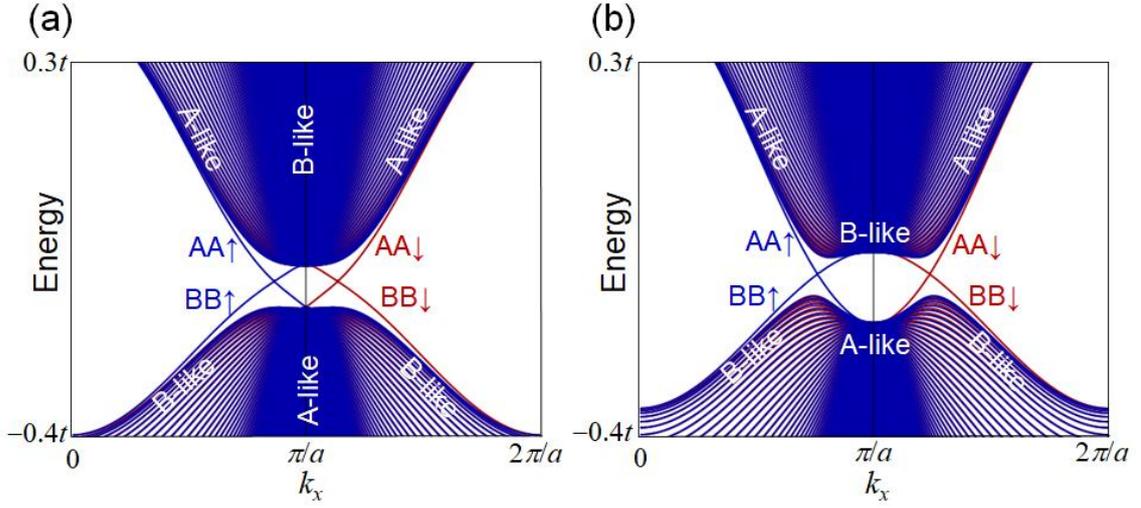

**FIG. 4.** (color online)

Energy dispersion of the nanoribbon of the $\tau$-type conductor parallel to the $x$-axis with a finite SOC of $\lambda = 0.01t$. (a) The isotropic case with no strain ($t_1 \equiv t$, $t_{2A} = t_{2B} = 0.13\,t$, $t_3 = -0.07\,t$), and (b) the anisotropic case with finite uniaxial strain ($t_1 \equiv t$, $t_{2A} = 0.13\,t \times 1.2$, $t_{2B} = 0.13\,t \times 0.8$, $t_3 = -0.07\,t$). "A-like" ("B-like") indicates that most of the state consists of A-site (B-site) states. "AA ↑" and "AA ↓" ("BB ↑" and "BB ↓") indicate the up-spin ($s_z = +1$) and down-spin ($s_z = -1$) edge states located around the AA-edge (BB-edge), respectively.

**Figure 5** (T. Osada)

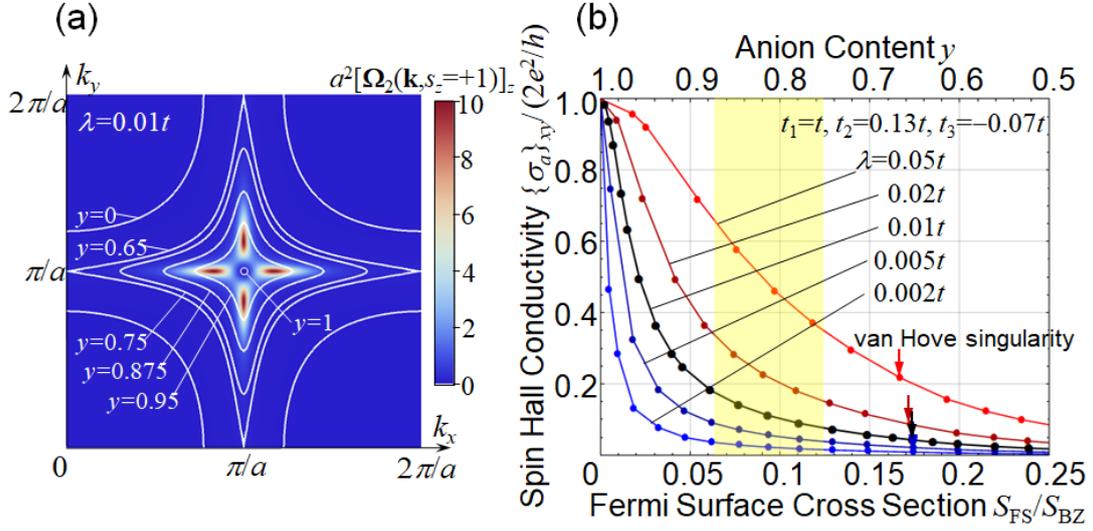

**FIG. 5.** (color online)

(a) Fermi surface and up-spin Berry curvature of the conduction band calculated for $\tau$-(EDO-$S,S$-DMEDT-TTF)$_2$(AuBr$_2$)$_{1+y}$ with a finite SOC of $\lambda = 0.01t$. White curves are Fermi surfaces for several anion contents. The color plot indicates the $z$-component of the Berry curvature. (b) Spin Hall conductivity as a function of the anion content, or the cross section of the Fermi surface for several SOC magnitudes. The shaded region shows the anion content in actual crystals. The Fermi level reaches the van Hove singularity at the filling indicated by arrows.

**Figure 6** (T. Osada)

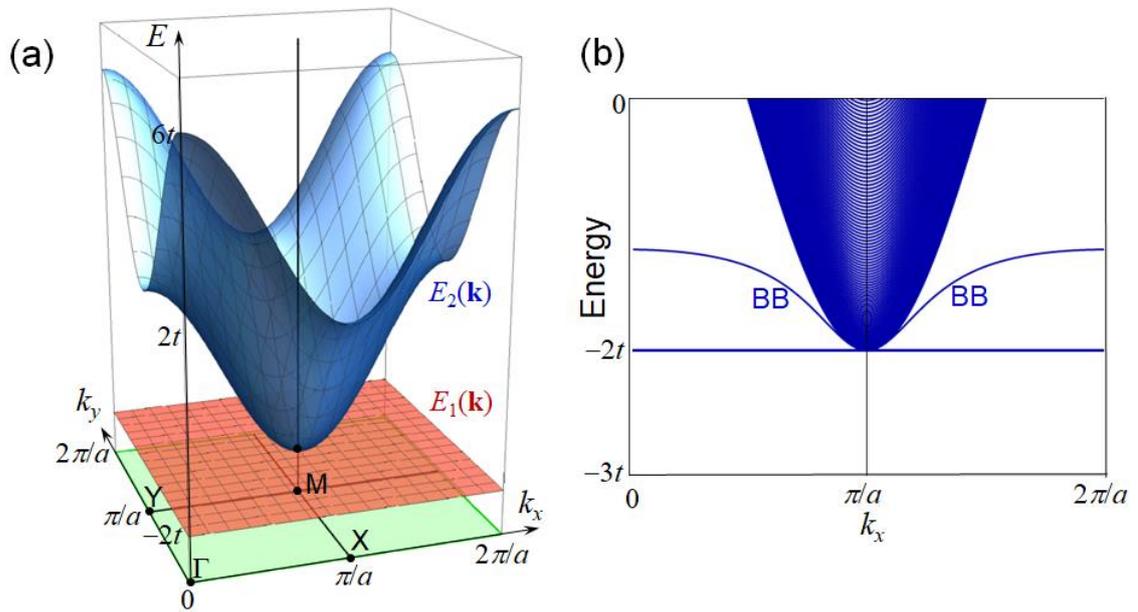

**FIG. 6.** (color online)

(a) Dispersion of the valence band $E_1(\mathbf{k})$ and the conduction band $E_2(\mathbf{k})$ of the Mielke lattice. (b) Energy dispersion of the Mielke lattice nanoribbon parallel to the *x*-axis with no SOC. An edge state labeled BB appears along the BB-edge of the nanoribbon.